\documentclass[twocolumn,floats,aps,showpacs,floatfix]{revtex4}
\usepackage{times,amsmath,graphicx,latexsym}
\begin{document}

\title{Vortex Fluctuations in the Critical Casimir Effect of Superfluid and Superconducting Films}

\author{Gary A. Williams},

\affiliation{Department of Physics and Astronomy, University of California, Los Angeles, CA 90095 USA}
\date{\today}

\begin{abstract}
Vortex-loop renormalization techniques are used to calculate the magnitude of the critical Casimir forces in superfluid films.  The force is found to become appreciable when size of the thermal vortex loops is comparable to the film thickness, and the results for $T<T_c$ are found to match very well with perturbative renormalization theories that have only been carried out for $T>T_c$. When applied to a high-$T_c$ superconducting film connected to a bulk sample, the Casimir force causes a voltage difference to appear between the film and bulk, and estimates show that this may be readily measurable.
\end{abstract}

\maketitle

Critical fluctuations in a finite-size superfluid lead to a free-energy difference between the finite-size system and the bulk.  If there is a connection between the two, such as between a saturated helium film and a bulk liquid reservoir, forces (known as Casimir forces, in analogy with size-limited electromagnetic fluctuations) will develop that lead to a thinning of the film in the vicinity of the superfluid transition, an effect which has been observed experimentally \cite{hallock,chan}.  Existing theories of the effect are incomplete:  perturbative $\epsilon$-expansion theories \cite{krech} are only able to calculate the force in the non-superfluid region $T>T_c$, and the maximum amplitude predicted for Dirichlet boundary conditions is about a factor of 50 times smaller than the observed maximum \cite{chan}.  The superfluid regime $T<T_c$ is apparently far more difficult for the perturbation theories, and such calculations have not yet been attempted.

We show here that vortex excitations \cite{cosmic,shenoy} are the source of the critical fluctuations giving rise to the critical Casimir force, and that vortex renormalization techniques provide a very simple means of calculating the force in the superfluid phase $T<T_c$.  The force becomes appreciable when the size of the thermally-excited vortex loops become comparable to the film thickness, and the results for periodic boundary conditions match very well with the perturbation theories at $T_c$.  When the loops become larger than the film thickness there is then a crossover to two-dimensional (2D) Kosterlitz-Thouless (KT) vortex pairs, and this leads to a prediction that the KT superfluid transition will take place at a temperature only slightly higher than the point where the Casimir force begins to be measurable: in helium films nearly the entire dip in thickness will occur in the normal state above $T_{KT}$.

We also propose that an analogous Casimir force should appear at the junction between a high-$T_c$ superconducting film and the bulk superconductor.  In this case the force will take the form of an electrical potential difference appearing between the film and bulk, due to a transfer of Cooper pairs from the film to the bulk that balances the Casimir energy difference.  Rough estimates show that this may be a readily measurable voltage (microvolts), and could be a useful probe of the high-$T_c$ superfluid transition.
\begin{figure}[t]
\begin{center}
\includegraphics[width=0.48\textwidth]{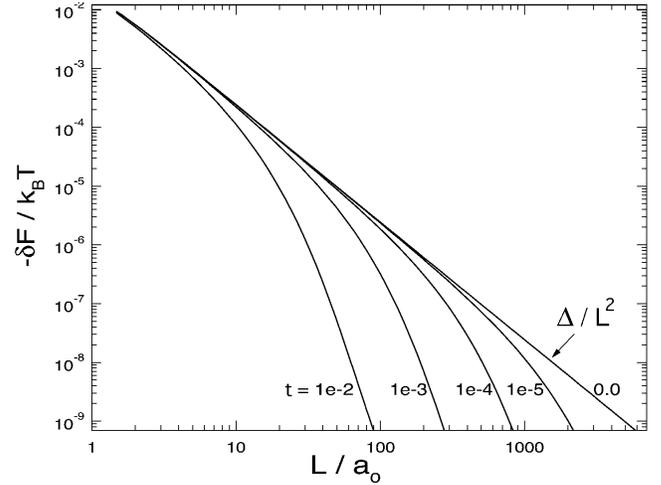}
\end{center}
\caption{Free-energy difference between bulk and film as a function of the film thickness L, for several reduced temperatures
$t=(T_c-T)/ T_c$.}
\label{fig1}\end{figure}

The difference in free energy per unit area between the film of thickness $L$ and the bulk is given by
$\delta F=L(f_b-f_f)$ where $f_b$ and $f_f$ are the free energies per unit volume of the bulk and film.  In the vortex-loop renormalization scheme these free energies \cite{shenoy} can be written as an integral over the average loop diameter $a$, and the difference is then
\begin{equation}
\label{eq1}
{\frac {\delta F}{k_BT}}=-L{ \frac {\pi}{a_o^3}}\int_{\beta L}^\infty  {\left( {\frac {a}{a_o}} \right)^{2}\exp \left( {-U(a)/k_BT} \right){\frac {da}{a_o}}}
\end{equation}
where $a_o$ is the bare core diameter ( = 2.53 \AA \,\,for helium parameters \cite{jltp}), $U(a)$ is the renormalized loop energy \cite{cosmic},  and $\beta L$ is the  maximum loop size in the film;  a comparison \cite{jltp} with a finite-size path-integral Monte Carlo simulation gave $\beta$ = 0.75.  In Ref.~\cite{cosmic} it was noted that from the scaling relation for the superfluid density the Boltzmann factor can be written in the form 
\begin{equation}
\label{eq2}
\exp \left( {-U(a)/k_BT} \right)={\frac {3\,a_o}{4\pi ^3}}\left( {\frac {a}{a_o}} \right)^{-6}{ \frac {\partial}{\partial a}}\left( {\frac {1}{K_r}} \right) 
\end{equation}
where $K_r=\hbar ^2\rho _sa_o/m^2k_BT$ is the dimensionless superfluid density.  Figure 1 shows an evaluation of Eq.~\ref{eq1} as a function of the film thickness $L$, using the loop recursion relations  in Ref.~\cite{cosmic}, for several reduced temperatures near $T_c$.  Very close to T$_c$ there is a crossover from exponential to algebraic decay in $L$, since at $T_c$ the  asymptotically  exact solution of the recursion relation is
$K_r=D_o\left( {{{a_o} \mathord{\left/ {\vphantom {{a_o} a}} \right. \kern-\nulldelimiterspace} a}} \right)$
where $D_o$ = 0.3875 is a universal constant \cite{cosmic,shenoy}.  Inserting this into Eqs.~\ref{eq1} and \ref{eq2} gives 
${\delta F} / {k_BT}={\Delta / L^2}$ precisely at T$_c$, where $\Delta ={{-1} / ({4\pi ^2 D_o \beta ^3})}$.
\begin{figure}[t]
\begin{center}
\includegraphics[width=0.48\textwidth]{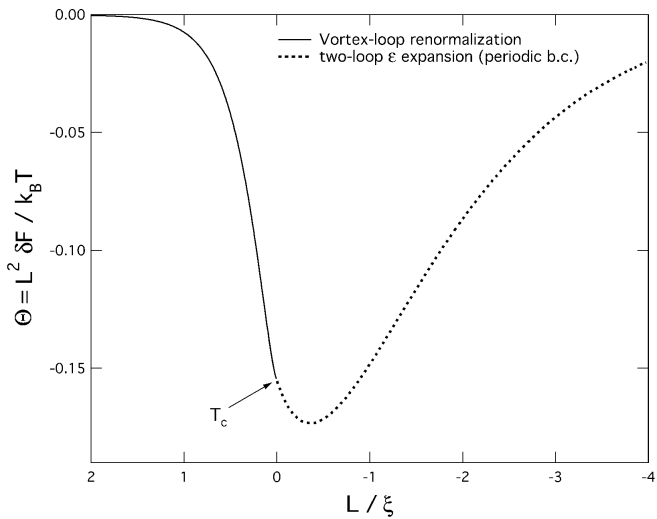}
\end{center}
\caption{Free energy scaling function versus $L / \xi$, where $\xi$ is the bulk correlation length, taken to be positive for T $<$ T$_c$ and negative for T $>$ T$_c$.}
\label{fig2}\end{figure}
This is exactly the form initially predicted by Fisher and DeGennes \cite{fisher} from scaling arguments, where the universal constant $\Delta$ is known as the Casimir amplitude.  With $\beta$ = 0.75 we find $\Delta$ = -0.155, in very reasonable agreement with the value $\Delta$ = -0.20 found in the $\epsilon$-expansion results for periodic boundary conditions \cite{krech}.  As in Ref.~\cite{jltp} our calculation is equivalent to periodic boundary conditions, since we assume the superfluid density is a constant across the film, with no variation at the wall or free surface.
Figure 2 shows the free energy scaling function 
$\Theta ={{L^2\,\delta F} \mathord{\left/ {\vphantom {{L^2\,\delta F} {k_BT}}} \right. \kern-\nulldelimiterspace} {k_BT}}$, plotted as a function of the scaling variable ${L \mathord{\left/ {\vphantom {L \xi }} \right. \kern-\nulldelimiterspace} \xi }$ where $\xi ={{a_o} \mathord{\left/ {\vphantom {{a_o} {K_r}}} \right. \kern-\nulldelimiterspace} {K_r}}$ is the bulk correlation length, the size of the largest loops being thermally excited.  The solid line is the vortex-loop result from evaluating Eq.~\ref{eq1}, and where for plotting purposes we have taken  $\xi$ positive for $T< T_c$.  When the maximum loop size becomes comparable to the film thickness the free energy difference decreases rapidly, with a finite slope at $T_c$ where ${L \mathord{\left/ {\vphantom {L \xi }} \right. \kern-\nulldelimiterspace} \xi }$ = 0. The dotted curve is the $\epsilon$-expansion result \cite{krech} for periodic boundary conditions, normalized by the ratio of the Casimir amplitudes, 0.155/0.20.  It is clear there is good agreement between the two calculations, with the finite slope at $T_c$ matching well.

The Casimir force $K_c$ leading to the film thinning is the derivative of the free energy difference, and is most conveniently written \cite{krech} in terms of a scaling function $\vartheta$,
$K_c=-{{\partial \;\delta F} / {\partial L}}=k_BT{\vartheta / {L^3}}$.
Figure 3 shows the results for the scaling function, which can be extracted from the experimental data as in Ref.~\cite{chan}.  This cannot be directly compared with the experimental results, however, since the superfluid density in a real film falls to zero at the boundaries, and hence Dirichlet boundary conditions rather than periodic should be applied.  The depression of the superfluid density at the surfaces also has the effect of shifting the transition temperature downward from the bulk critical temperature $T_{\lambda}$ by an amount dependent on $L$, $T_c$ = $T_c(L)$.  It should be possible in further work with the vortex-loop theory to account for the depressed superfluid density by calculating the excess loop density near a wall \cite{3d2d}. 

A further effect which must be taken into account is the crossover from 3D to the 2D KT transition.  In the above calculation the iterations in the film have been stopped when the loop size reaches the film thickness, but actually at that point the loops intercept the boundaries and turn into vortex pairs at longer length scales.   Adding these excitations to the above Casimir calculation is easily carried out \cite{3d2d} by matching the loop recursion relations to the Kosterlitz recursion relations\cite{kosterlitz} for the vortex pairs at the crossover length $\beta L$.  The areal superfluid density $\sigma_s$ is related to $\rho_s$ resulting from the loop recursion relations by $\sigma_s$ = $\rho_s L$, and the fugacity of the pairs is set equal to the loop fugacity at that length scale.  The core diameter of the pairs is $a_c$, the effective core diameter of the loops at the crossover\cite{cosmic,shenoy}, which is proportional to the correlation length, and hence is of the order of the film thickness $L$.  

\begin{figure}[t]
\begin{center}
\includegraphics[width=0.48\textwidth]{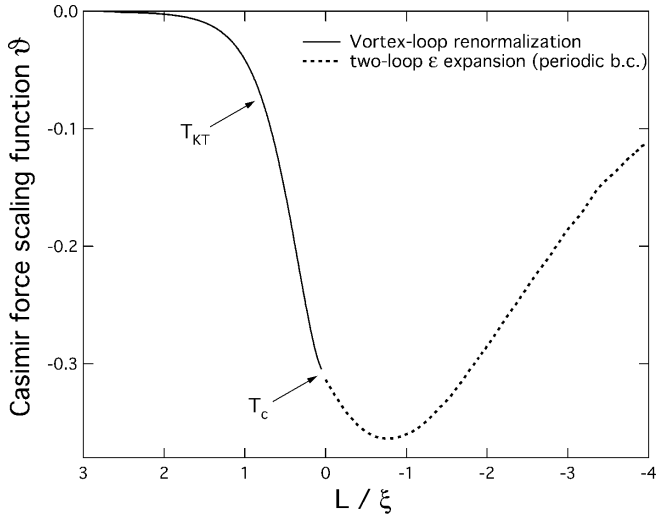}
\end{center}
\caption{Casimir force scaling function $\vartheta$ versus $L / \xi$.}
\label{fig3}\end{figure}

With these inputs the KT recursion relations are then iterated to macroscopic length scales.  The superfluid density jumps to zero at the temperature $T_{KT}$  indicated in Fig.~\ref{fig3}.  $T_{KT}$ is a function of the scaling variable $L / \xi$, as shown previously in Ref.~\cite{3d2d} where agreement was found with finite-thickness scaling of the KT transition\cite{ahns}.  Since the film thinning also begins to occur when the correlation length is comparable to the film thickness, it is not surprising that the KT transition occurs close to the onset of the film thinning, with nearly all of the film thinning occurring in the normal state above $T_{KT}$.  By also iterating the Kosterlitz free energy along with the recusion relations, the 2D contribution to the Casimir forces can be evaluated.  We find that for thick films ($L / a_0 \sim$ 100) this is negligible compared to the contribution from the loops, since the Kosterlitz free energy for the vortex pairs is proportional to $a_c^{-2}\approx L^{-2}$, reducing the pair free energy per unit area by a factor of  $(a_o / L)^{2}$
compared to that of the loops.  An important experimental check of these predictions for the KT onset point will be to measure the superfluid density simultaneously with the film thinning and check whether $T_{KT}$ is coincident with the onset of thinning.  Although again the present results apply only for the case of periodic boundary conditions, it is likely that both $T_c$ and $T_{KT}$ scale in just the same way for Dirichlet boundary conditions.

A similar Casimir force should also be present near the phase transition of a high-$T_c$ superconducting thin film connected to a bulk sample of the same material.  There is now considerable theoretical and experimental evidence that the high-$T_c$ transition is also a vortex-loop transition entirely similar to that of helium \cite{cosmic,shenoyhtc,htc}.  To give a concrete example of this we show in Figure 4 a fit of the vortex-fluctuation theory of 
Shenoy and Chattopadhyay \cite{shenoyhtc} to recent experimental data \cite{bscco} for the superfluid fraction of a Bi$_2$Sr$_2$CaCuO$_2$ (BSCCO) epitaxial film of thickness 610 \AA .  This film (labeled curve B in Ref.~\cite{bscco}) is slightly overdoped, with a $T_c$ of 84.9 K.  At low temperatures vortices are not thermally excited, and the only excitations that affect the superfluid density are the nodal quasiparticles.  Fitting to the data in the range from 5 to 20 K gives a "bare" superfluid fraction $\rho^o_s / \rho = 1 - A T^2$ with $A$ = 7.5$\times$10$^{-4}$ K$^{-2}$, and this provides a good description up to about 60 K where the vortices take over.  The $T^2$ decrease of the superfluid fraction is the form expected for d-wave quasiparticles in the dirty limit \cite{qp}.

\begin{figure}[t]
\begin{center}
\includegraphics[width=0.48\textwidth]{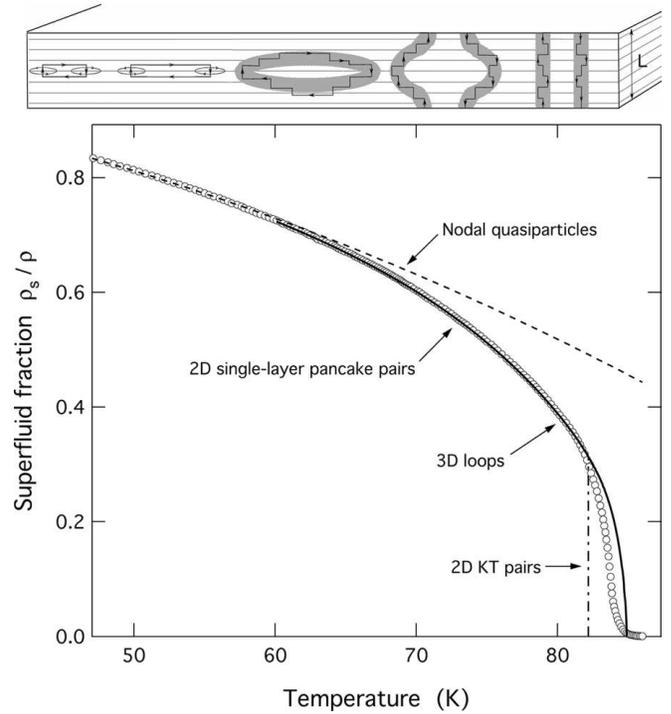}
\end{center}
\caption{Superfluid fraction in a BSCCO film, with data from Ref.~\cite{bscco}.  The dashed curve is the background from the nodal quasiparticles, the solid curve that from single-layer vortex pairs and loops, and the dash-dot line indicates the location of the KT jump.  The schematic figure above the plot indicates the progression with temperature of the 2D single-layer pairs, 3D anisotropic loops, and the crossover back to 2D vortex pairs.}
\label{fig4}\end{figure}

Above 60 K the initial vortex excitations are single-layer pancake-antipancake pairs\cite{shenoyhtc}.  These are particularly important in BSCCO due its strong anisotropy, with an anisotropy factor at small length scales $\gamma_o$ = $\xi_{\|} /  \xi_{\bot}$ = 50 \cite{gamma}, where $\xi_{\|}( \approx$ 25 \AA ) and $\xi_{\bot}$ are the correlation lengths in the directions parallel and perpendicular to the CuO planes.  Modified KT recursion relations that include the linear potential from the Josephson cores are iterated from an initial pair separation $a_o$ = $\xi_{\|} $ as in Ref.~\cite{shenoyhtc}, except that the bare superfluid fraction is taken to be that of the nodal quasiparticles above.  The only fitting parameter is the core energy of a pancake vortex; the fit to the data gives a value of the core energy over the bare superfluid density $E_c$ / $K_o$ = 1.5, in units $k_BT$.  This is quite comparable to the value of 2.2 found in helium films\cite{kotsubo}.  

At the length scale $r_o$ = $\gamma_o a_o$ there is then a crossover to vortex loops, which initially are quite elliptical due to the strong anisotropy, as shown in Fig.~\ref{fig4}.  The starting fugacity and superfluid density of the loops are matched to those of the 2D pairs at $r_o$, and for a bulk sample the 3D loop recursion relations are then iterated to scales greater than the coherence length, where the superfluid density becomes constant.  The result for a bulk sample is shown as the solid line in Fig.~\ref{fig4}, in excellent agreement with the film data at least to within a few degrees of $T_c$.  Beginning about 2 K from  $T_c$ the recursion relation for the anisotropy factor\cite{shenoyhtc} shows that it scales rapidly towards one, so that purely circular loops are excited at large length scales.  In this limit where $\rho_s / \rho  <$ 0.1 the superfluid exponent becomes\cite{shenoy} $\nu$ = 0.6717, the XY model value.  The temperature range where XY critical behavior can be observed in BSCCO is considerably smaller than in other high-$T_c$ materials because of its large anisotropy factor.

\begin{figure}[t]
\begin{center}
\includegraphics[width=0.35\textwidth]{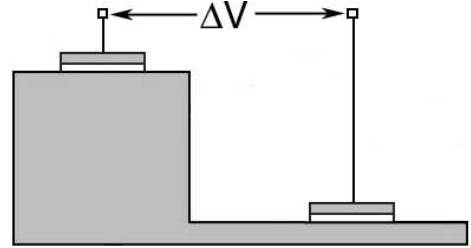}
\end{center}
\caption{Schematic of an experiment to measure the voltage difference between film and bulk, using two Josephson junctions.}
\label{fig5}\end{figure}

The reason the film data falls away from the predicted line for the bulk is likely due to a crossover to  the 2D KT transition, which will occur when the loop size becomes comparable to the film thickness, as discussed above for helium films.  For this case the superfluid density and fugacity are matched at the scale 0.5 $L$, to better account for the depressed superfluid density at the surfaces in the experiments.  The 2D KT recursion relations are then iterated to larger length scales.  The KT transition occurs quickly after the crossover;  the dash-dot line in Fig.~\ref{fig4} shows the universal jump of the superfluid density to zero, which coincides well with the deviation of the film data from the bulk curve.  The experimental data is not expected to show a sharp jump to zero, since it is taken at a rather high frequency of 80 kHz.  It is well known that this leads to a finite-frequency broadening \cite{ahns,minn} quite similar to the behavior seen in the data here.

The critical Casimir free energy difference between a BSCCO film and  bulk can now be calculated in just the same manner as for helium films.  The quasiparticle and single-layer pair contributions are the same in the film and bulk and cancel, and the KT pairs give a negligible contribution.  The difference is entirely from the loop size being cut off by the film thickness, and the result for periodic boundary conditions is essentially the same as Fig.~\ref{fig2}.  The free energy difference leads to a transfer from the film to the bulk of $\delta N$ Cooper pairs per unit area of the film, and this gives rise to an electrical potential difference $\delta V$ that appears at the interface between the film and bulk.  Since the total electrochemical potential $\mu + eV$ must be a constant throughout the system, the free energy difference is proportional to the electrical potential difference $\delta V = V_b-V_f$,
\begin{equation}
\delta F=(\mu _b-\mu _f)\;\delta N=-e\;\delta V \;\delta N
\end{equation}
But the number of Cooper pairs that transfer must also be proportional to the voltage difference\cite{pethick}, 
\begin{equation}
\delta N=-2\,N(E_F)\;L\,(\mu _b-\mu _f)\;=2\,N(E_F)\;L\;e\;\delta V
\end{equation}
where N(E$_F$) is the density of single-spin electronic states at the Fermi surface.  Combining these gives a prediction for the voltage difference 
$\delta V=\sqrt {-\delta F/2\,e^2N(E_F)\;L}$.
This can be crudely estimated for a BSCCO film 600 \AA \,thick by taking a maximum ${\delta F} / {k_BT_c}\approx {\Delta / L^2}$ with $\Delta \approx$ -0.15 and $T_c \approx$ 85 K, N(E$_F$) $\approx$ 1$\times$10$^{46}$/ Jm$^3$, yielding a maximum $\delta V \approx$ 30 $\mu$V.  This is the estimate for periodic boundary conditions and not Dirichlet, but we note that the experimental results  in helium films appear to give an even larger value of $\Delta$ than used here.

It may be possible to measure this voltage with the schematic experiment shown in Fig.~\ref{fig5}, where two high-$T_c$ Josephson junction contacts equilibrate with the chemical potential in the film and bulk, and the voltage difference $\delta V$ across them can be measured with a low-$T_c$ SQUID voltmeter.  This technique has been used to measure voltage differences in low-$T_c$ superconductors\cite{clarke} to a resolution of better than nanovolts, and it should be possible to apply the same technique to the Casimir voltage measurement.  The electrodes of the two Josephson junctions need to have a higher $T_c$ than the bulk or film;  they can either be a different high-$T_c$ material, or the same material but closer to optimal doping.  Observation of this voltage would give a sensitive new probe of the high-$T_c$ phase transition, and would allow measurements of the perpendicular coherence length near the transition, since the voltage is only appreciable when the coherence length equals the film thickness.
In summary, we have shown that the critical Casimir effect in superfluid and superconducting films results from the finite-size limitation of vortex loop fluctuations in the film.  This is the first calculation of the effect in the superfluid phase;  the ease of evaluating Eq.~\ref{eq1} using the loop renormalization should be contrasted with the extreme complexity of traditional perturbative methods that prevent the calculations from being carried out below $T_c$.   For periodic boundary conditions the loop result matches with the perturbation theories at  $T_c$, but a full comparison with experiments will require extension of the theory to Dirichlet boundary conditions.

We thank K. Osborn, D. Van Harlingen, R. Simmonds, R. Packard, M. Chan, and M. Krech for useful discussions and correspondence.  The hospitality of the Physics Department at UC Berkeley is gratefully acknowledged, where this work was begun.
Work supported by the National Science Foundation, DMR 01-31111.

\end{document}